\documentclass[12pt]{iopart}

\usepackage{iopams}  
\usepackage{graphicx}
\begin{document}
	
	\title{I-mode investigation on the Experimental Advanced Superconducting Tokamak}
	
	\author{X. Feng$^1$, A.D. Liu$^1$, C. Zhou$^1$, Z.X. Liu$^1$, M.Y. Wang$^2$, G. Zhuang$^1$, X.L. Zou$^3$ T.B. Wang$^3$$^4$$^5$, Y.Z. Zhang$^6$, J.L. Xie$^1$, H.Q. Liu$^7$, T. Zhang$^7$, Y. Liu$^7$, Y.M. Duan$^7$, L.Q. Hu$^7$, G.H. Hu$^7$, D.F. Kong$^7$, S.X. Wang$^7$, H.L. Zhao$^7$, Y.Y. Li$^7$, L.M. Shao$^7$, T.Y. Xia$^7$, W.X. Ding$^1$, T. Lan$^1$, H. Li$^1$, W.Z. Mao$^1$, W.D. Liu$^1$, X. Gao$^7$, J.G. Li$^7$, S.B. Zhang$^7$, X.H. Zhang$^8$, Z.Y. Liu$^1$, C.M. Qu$^1$, S. Zhang$^1$, J. Zhang$^1$, J.X. Ji$^1$, H.R. Fan$^1$, and X.M. Zhong$^1$}
	\address{$^1$ KTX Laboratory and Department of Engineering and Applied Physics, University of Science and Technology of China, Anhui Hefei 230026, China}
	\address{$^2$ Department of Physics, Nanchang University, Nanchang 330031, China}
	\address{$^3$ Institute for Magnetic Fusion Research, CEA, F-13115 Saint-Paul-lez-Durance, France}
	\address{$^4$ Southwestern Institute for Physics, CNNC, C-610200 Chengdu, China}
	\address{$^5$ Department of Applied Physics, Ghent University, B-9000 Ghent, Belgium}
	\address{$^6$ Center for Magnetic Fusion Theory, CAS, Hefei, Anhui 230026, China}
	\address{$^7$ Institute of Plasma Physics, Chinese Academy of Sciences, Anhui Hefei 230031, China}
	\address{$^8$School of Computer and Information, Hefei University of Technology, Hefei, Anhui 230009, China}
	
	\ead{zhouchu@ustc.edu.cn}
	\ead{zxliu316@ustc.edu.cn}
	\vspace{10pt}
	\begin{indented}
		\item[]May 2019
	\end{indented}
	
	\begin{abstract}
		By analyzing large quantities of discharges in the unfavorable ion $ \vec B\times \nabla B $ drift direction, the I-mode operation has been confirmed in EAST  tokamak. During the L-mode to I-mode transition, the energy confinement has a prominent improvement by the formation of a high- temperature edge pedestal, while the particle confinement remains almost identical to that in the L-mode. Similar with the I-mode observation on other devices, the $ E_r $ profiles obtained by the eight-channel Doppler backscattering system (DBS8)\cite{J.Q.Hu} show a deeper edge $ E_r $ well in the I-mode than that in the L-mode. And a weak coherent mode (WCM) with the frequency range of 40-150 kHz is observed at the edge plasma with the radial extend of about 2-3 cm. WCM could be observed in both density fluctuation and radial electric field fluctuation, and the bicoherence analyses showed significant couplings between WCM and high frequency turbulence, implying that the $ E_r $ fluctuation and the caused flow shear from WCM should play an important role during I-mode. In addition, a low-frequency oscillation with a frequency range of 5-10 kHz is always accompanied with WCM, where GAM intensity is decreased or disappeared. Many evidences show that the a low-frequency oscillation may be a novel kind of limited cycle oscillation but further investigations are needed to explain the new properties such as the harmonics and obvious magnetical perturbations.
	\end{abstract}
	\\
	\section{Introduction}
	\setlength{\parindent}{2em}
	\indent
	As the H-mode with type-I edge localized modes (ELM) can generate unacceptable heat loads to the divertor and plasma-facing components, the I-mode is being explored as an alternative operating mode for future devices. The I-mode simultaneously features high-energy confinement, such as the H-mode, and low particle confinement, such as the L-mode without ELMs, and this type of quiet pedestal can protect the divertor target from ELM heat flux with an acceptable plasma energy confinement.

	The I-mode was originally observed as an energy confinement improved mode in 1990s\cite{F.Ryter-2}; currently, it has been widely investigated in Alcator C-Mod\cite{D.G.Whyte,A.E.Hubbard-4,R.M.McDermott,C.Theiler,A.E.Hubbard-3,I.Cziegler,A.E.White-1,A.E.White-3}, AUG\cite{F.Ryter,E.Viezzer,P.Manz-2,P.Manz-1,T.Happel-1,T.Happel-2}, and DIII-D\cite{A.Marinoni}. In unfavorable configurations\cite{A.E.Hubbard-1}, the L-mode to I-mode transition occurs when the heating power exceeds a certain threshold\cite{A.E.Hubbard-2}, and an edge energy transport barrier\cite{A.E.Hubbard-3} is constructed accompanied with the weak coherent mode (WCM)\cite{P.Manz-2,I.Cziegler,A.E.White-1,A.E.White-3,A.Marinoni} at the steepest temperature gradient region. In the I-mode operation, an $ E_r $ well\cite{R.M.McDermott,C.Theiler,A.Marinoni,E.Viezzer} forms at the edge of the plasma, and a stable temperature pedestal is simultaneously established\cite{J.R.Walk,J.W.Hughes}. The poor particle confinement in the I-mode also contributes to a low core impurity level and sustains the operation performance.\cite{J.E.Rice,J.E.Rice-2}

	In this paper, the typical I-mode investigation in EAST is presented for the first time. In 2014, the upper divertor was upgraded to full tungsten, while the lower divertor was conserved to be carbon.\cite{Z.b.Zhou} Consequently, all high- performance discharges were in the upper single null shape to reduce the carbon impurities entering the main plasma.\cite{L.Zhang} Thus, operations most frequently have a clockwise toroidal field, such that the ion $ \vec B\times \nabla B $ drift directs away from the active X point, which provides a large amount of discharges for the I-mode investigation. After numerous equilibrium parameters and fluctuation signals have been analyzed, the I-mode operation in EAST is confirmed. L-mode to I-mode transition can be triggered by all types of heating methods, and no heating preference is observed, which is consistent with the results in other tokamaks\cite{F.Ryter,A.E.Hubbard-1}. In EAST, the WCM with the frequency of 40-150 kHz is observed in both density fluctuation and $ E_r $ fluctuation at the plasma edge using the eight-channel Doppler backscattering system (DBS8) in the I-mode operation, and a low frequency oscillation (5-10 kHz) is also observed when the WCM is present. The geodesic acoustic model  (GAM) may disappear at the location where the WCM is most significant, while it can still exist at other locations in I-mode on EAST.

	The paper is organized as follows. In section 2, the experimental conditions and eight-channel Doppler backscattering system for the fluctuation investigation are presented. The identification of the main features of the I-mode, including the density, temperature and radial electric field profiles, and the WCM are shown in section 3. The nonlinear interaction among the WCM, the low-frequency oscillation and background turbulence by a bicoherence analysis is also included in section 3. The last section is the summary.
	\section{Experimental conditions and diagnostics}
	All experiments were performed on the Experimental Advanced Superconducting Tokamak (EAST)\cite{XianzuGONG,J.Li,BaonianWan,Z.X.Liu-3}, which is the first full superconducting divertor tokamak with $ R_0 \sim $ 1.88 m, $ a\sim $0.45 m, $ I_p\leq$1.0 MA, $ B_T\leq $ 3.5 T. And the ITER-like configuration of EAST makes this work more meaningful. The EAST features complete RF wave heating, including lower hybrid wave heating (LHW), ion cyclotron resonance heating (ICRF), electron cyclotron resonance heating (ECRH), and two neutral beam injection (NBI) systems, one in the co-current direction while the other is in the counter.

	The turbulence measurements are achieved by the DBS8 system,\cite{J.Q.Hu} and it can simultaneously launch eight different frequencies (55, 57.5, 60, 62.5, 67.5, 70, 72.5 and 75 GHz) into the plasma. Because the probing beam is obliquely launched to the cutoff layer, the Doppler shift caused by the movement of the turbulence at the cutoff layer $ f_d $ can be written as $ f_d = u_\perp k_\perp/2\pi $, where $ u_\perp $ is the perpendicular velocity, and $ k_\perp $ is  the perpendicular wavenumber at the cutoff layer. In the laboratory frame, the perpendicular velocity is the sum of the turbulence phase velocity and background $ E \times B $ velocity, which is expressed as $ u_\perp = v_{phase} +v_{E\times B} $. In many cases, at the plasma edge, $ v_{phase} $ is much smaller than $ v_{E\times B} $, and we can calculate the radial electric field as $ E_r\sim u_\perp B $.\cite{G.D.Conway,G.D.Conway2}

	The backscattered signal is obtained as the in-phase ($ I=Acos \phi $) and quadrature ($ Q=Asin \phi $) of the I/Q mixer; this complex signal is written as $ Ae^{i\phi}= I + iQ $. Then, the phase can be calculated as $ \phi(t) = arctan[Q(t)/I(t)] $, and the signal amplitude is $ A(t) =\sqrt{I^2 (t) + Q^2 (t)} $, where $ A $ corresponds to the density fluctuation level and the phase change rate $ d\phi $ corresponds to the perpendicular velocity of the cutoff layer (also to the radial electric field with the relationship of $ u_\perp\sim E_r/B $).\cite{G.D.Conway2}  Two different methods are commonly used to extract the Doppler shift from the DBS signal: the center of gravity of the complex amplitude signal and the phase derivative method. In this study, we choose the phase derivative method to directly obtain the Doppler shift from the change in phase signal: $ f_d = d\phi/dt $\cite{G.D.Conway2}, since it is the simplest one with the best time resolution.

	DBS has been widely used for turbulence measurements with specific wavenumbers ($ 4-15\;cm^{-1} $ for the system in this article). For the WCM, which is a type of larger-scale fluctuation (wavenumber smaller than 2 $ cm^{-1} $), can not be directly observed in the backscattered signal because its wavenumber is too small. However, the WCM can modulate the cutoff layer,\cite{P.Manz-2} and the DBS signal is sensitive to the oscillation caused by this modulation. Thus, WCM features such as the frequency can also be observed in DBS signals.
	\section{ Experimental results and analysis}
	\begin{figure*}
		\includegraphics[width=1\linewidth]{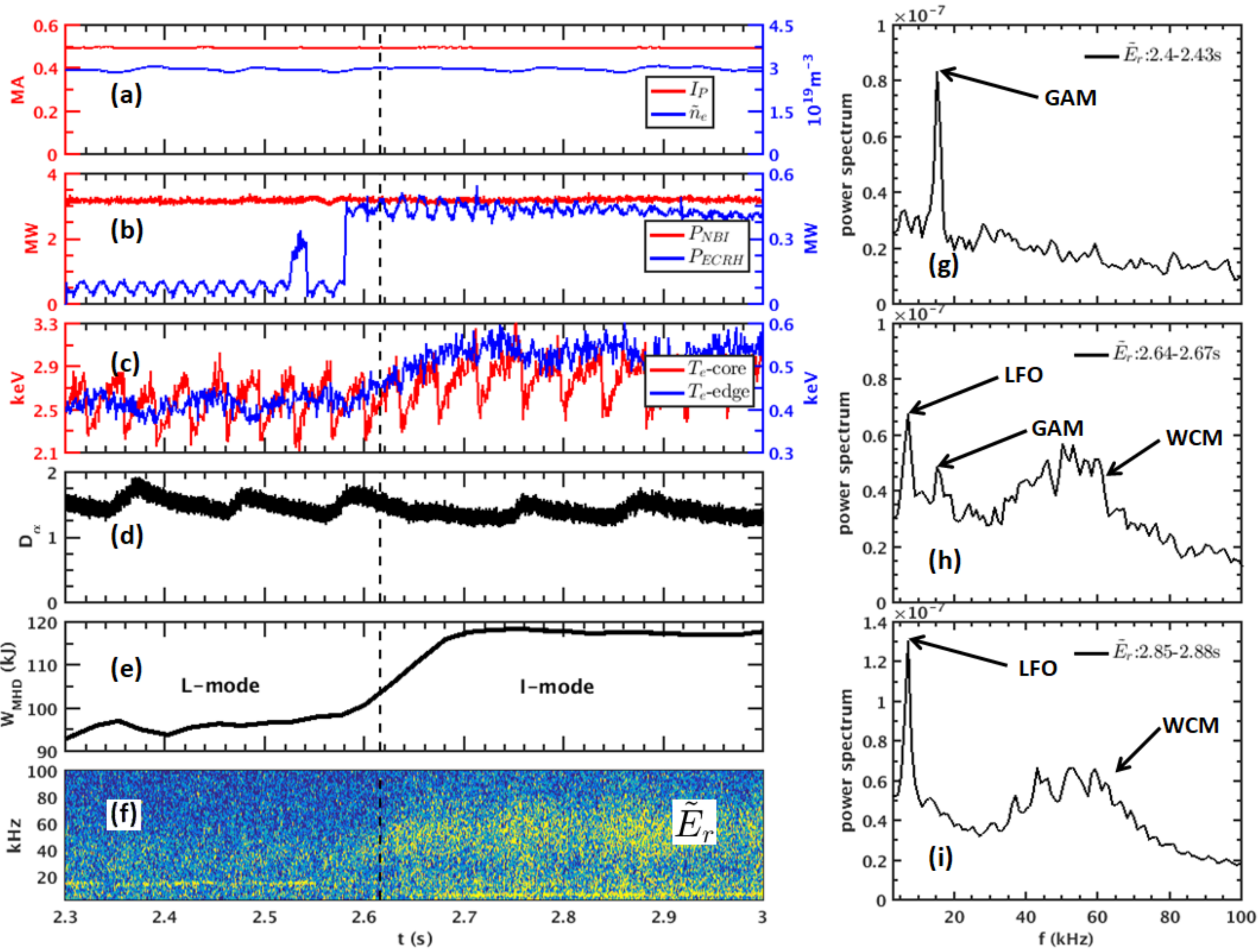}
		\caption{\label{fig:a} Example of upper single-null $  B_\phi $=2.23 T, $ q_{95}\sim5 $ I-mode discharge with L-mode, I-mode phases indicated. (a) Plasma current and line-averaged density $ \bar{n}_e$; (b) NBI and ECRH heating power; (c) $ T_e $ from ECE in core and edge; (d) $ D_\alpha $ recycling; (e) stored  energy $ W_{MHD} $; (f) WCM in the $ E_r $ perturbation. (g), (h), and (i) are power spectrum of the $ E_r $ perturbation for three time slices from the same channel used in panel (f).}
	\end{figure*}
	\begin{figure*}
		\includegraphics[width=1\linewidth]{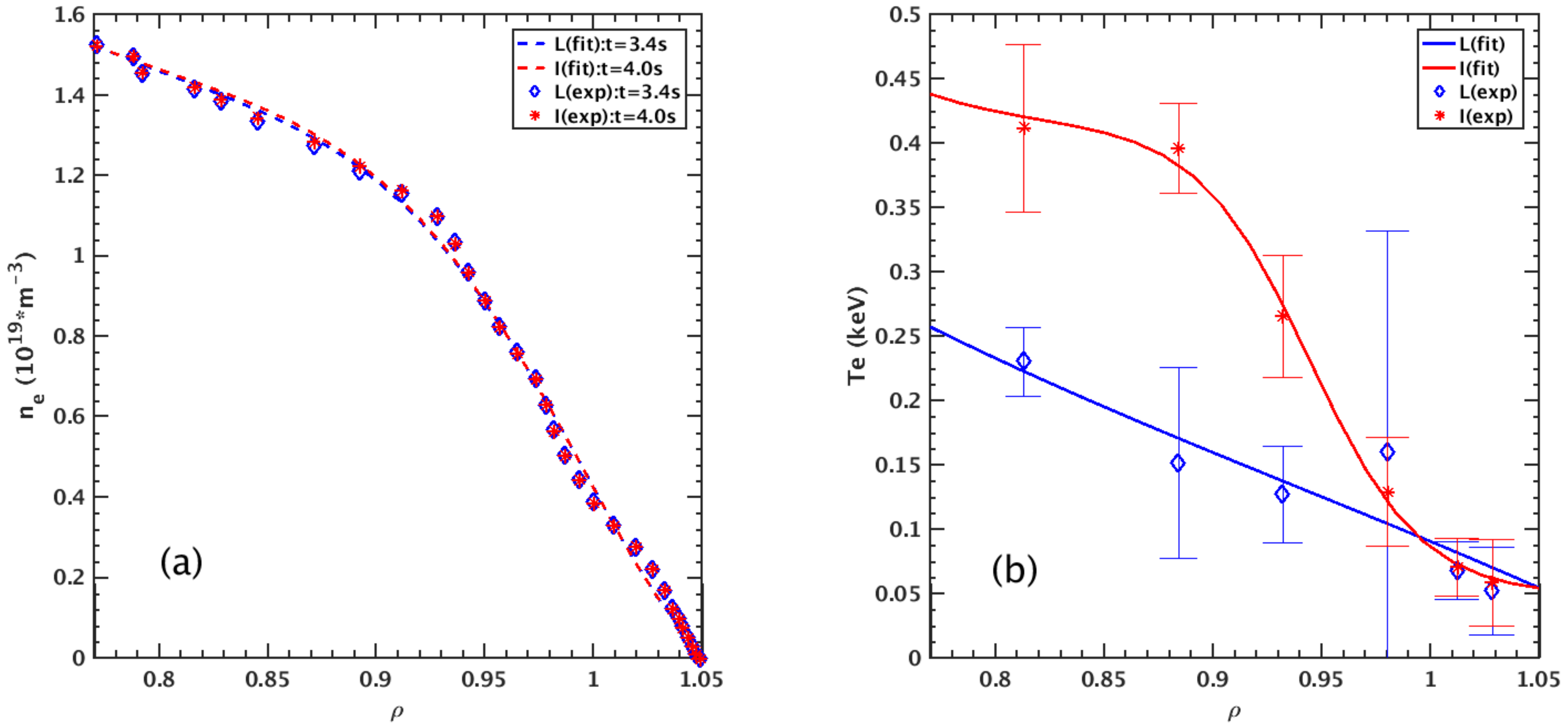}
		\caption{\label{fig:b} Density profile from reflectometer at 3.4 s in L-mode and 4.0 s in I-mode, time-averaged electron temperature profile from Thomson scattering (TS) in the L-mode and the I-mode at plasma edge in discharge 71078. And the solid lines and dashed lines are the fitting data and the diamonds and stars are the experimental data.
		}
	\end{figure*}

	\subsection{Identification of the I-mode}
	Currently, I-mode in EAST could be achieved under such parameter space with plasma currents of $ I_p\sim0.45 - 0.7\;MA $ and toroidal magnetic field $ B_\phi=2.26-2.49\; T $ under upper-single null (USN) geometries with an “unfavorable” ion $ \vec B\times \nabla B $ directed away from the active X-point\cite{A.E.Hubbard-4}. In particular, in discharges with heating-powers above 1.8 MW and line-averaged electron densities above $ 2.5*10^{19}\; m^{-3} $, the features of the I-mode become quite distinct. No heating method preference is observed, and the L-mode to I-mode transition can be triggered by various heating methods. A typical example of the I-mode discharge 69327 is shown in Fig.~\ref{fig:a}. This discharge is accompanied with sawtooth instability. The main heating methods are NBI and ECRH, and the power of NBI is constant at 3 MW, while ECRH turns on at 2.58 s with the power of 0.3 MW, as exhibited in Fig.~\ref{fig:a}(b). The plasma current is approximately 500 kA, and the line-averaged density from the HCN interferometer remains almost unchanged in the entire period as shown in Fig.~\ref{fig:a}(a). However, the core and edge electron temperature Te from Electron Cyclotron Emission (ECE) in Fig.~\ref{fig:a}(c) with stored energy $ W_{MHD} $ in Fig.~\ref{fig:a}(e) increase to a new plateau at approximately 2.6 s after the ECRH has turned on, which indicates the improvement of energy confinement. No density increase and no $ D_\alpha $ change in this transition were observed, which proves that the particle confinement remains similar to the L-mode. In the $ E_r $ fluctuation, the WCM is observed at the edge of the plasma after this transition as shown in Fig.~\ref{fig:a}(f), which further identifies the I-mode operation. Clearer exhibitions of fluctuation modes have been shown in panels (g) for L-mode, (h) for transition state, and (i) for I-mode, respectively. One notes that in this channel of DBS8 system ($ \rho\sim0.95 $) the GAM disappears gradually during the transition from L-mode to I-mode, while a low-frequency oscillation(LFO) and the WCM develop out. Using the energy confinement calculation formula $ \tau_{E}=W_{dia}/P_{loss}+\frac{\partial W_{dia}}{\partial t} $, a simple calculation shows that during this transition, the energy confinement time has a 7\% increase from 36.1 to 38.6 ms.

	To further confirm the I-mode, the density and temperature profiles were analyzed. The density profile from the reflectometer and electron temperature profile from Thomson scattering (TS) are shown in Fig.~\ref{fig:b} at plasma edge, where the L-mode are in blue and I-mode in red. Details on this discharge is shown in Fig.~\ref{fig:c}. It shows clearly that there is no significant difference in the density profile between in L-mode and I-mode as shown in Fig.~\ref{fig:b}(a). In the temperature profile, electron temperature grows a lot and a clear temperature pedestal forms at edge in I-mode region, where the data used here are time-averaged to improve signal-to-noise ratio of TS data. 
	\subsection{Features of the WCM}
	\begin{figure*}
		\includegraphics[width=1\linewidth]{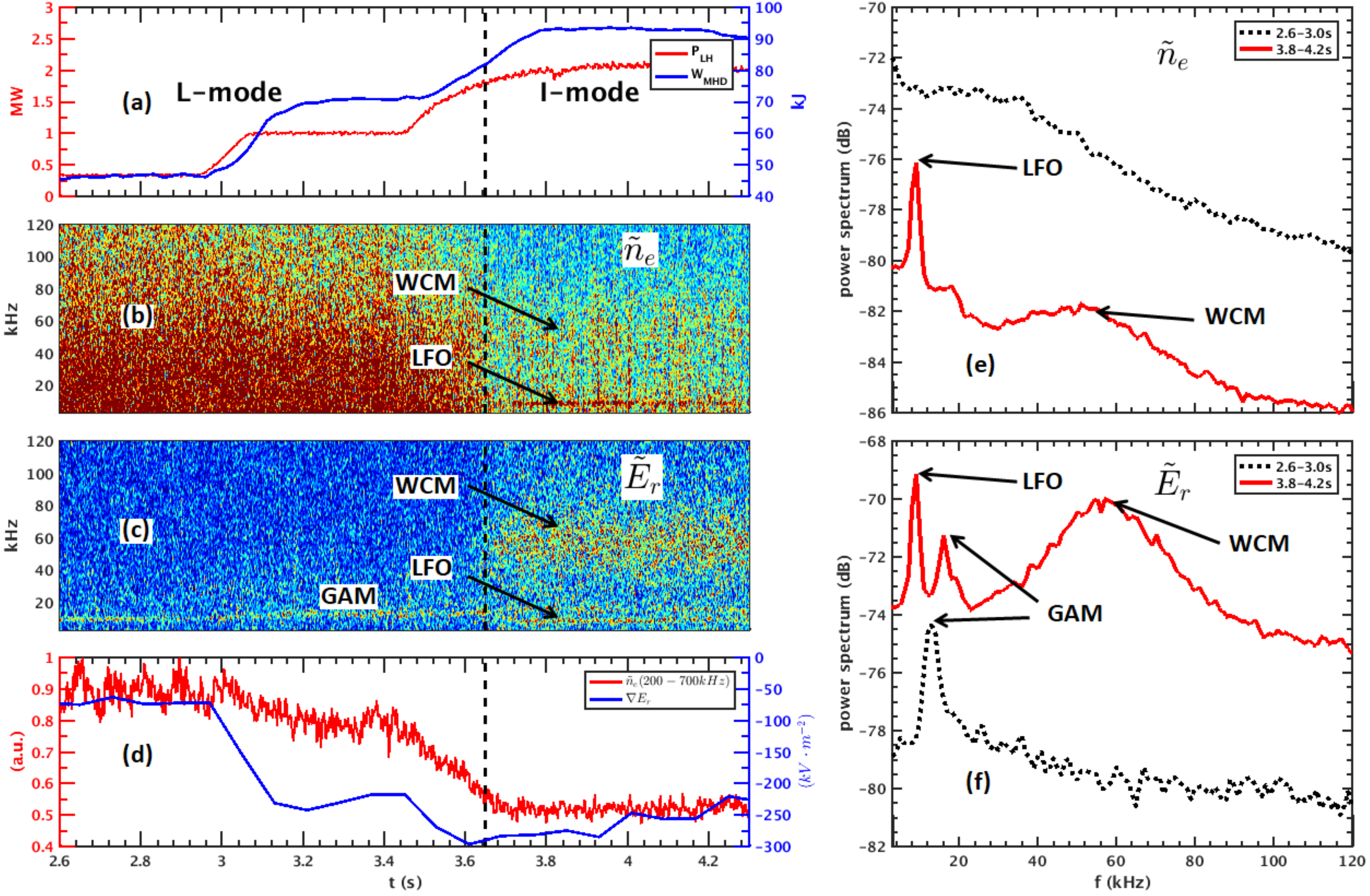}
		\caption{\label{fig:c} Example of the edge fluctuations from DBS8 at $ \rho\sim0.97 $ in discharge 71078. (a) is the LHW heating power in red and stored energy $W_{MHD}$ in blue. (b) and (c) are the spectrograms of the density perturbation and $ E_r $ perturbation from the L-mode to the I-mode, respectively. (d) is the evolutions of density perturbation amplitude in red with an integral range of 200-700 kHz and the shear of radial electric-field in blue. (e) and (f) are power spectrum of the density perturbation and $ E_r $ perturbation in both L-mode and I-mode. It should be mentioned that the red real line in panel (f) has been up-shifted by 3 dB to make peaks clearer.}
	\end{figure*}
	\begin{figure*}
		\includegraphics[width=1\linewidth]{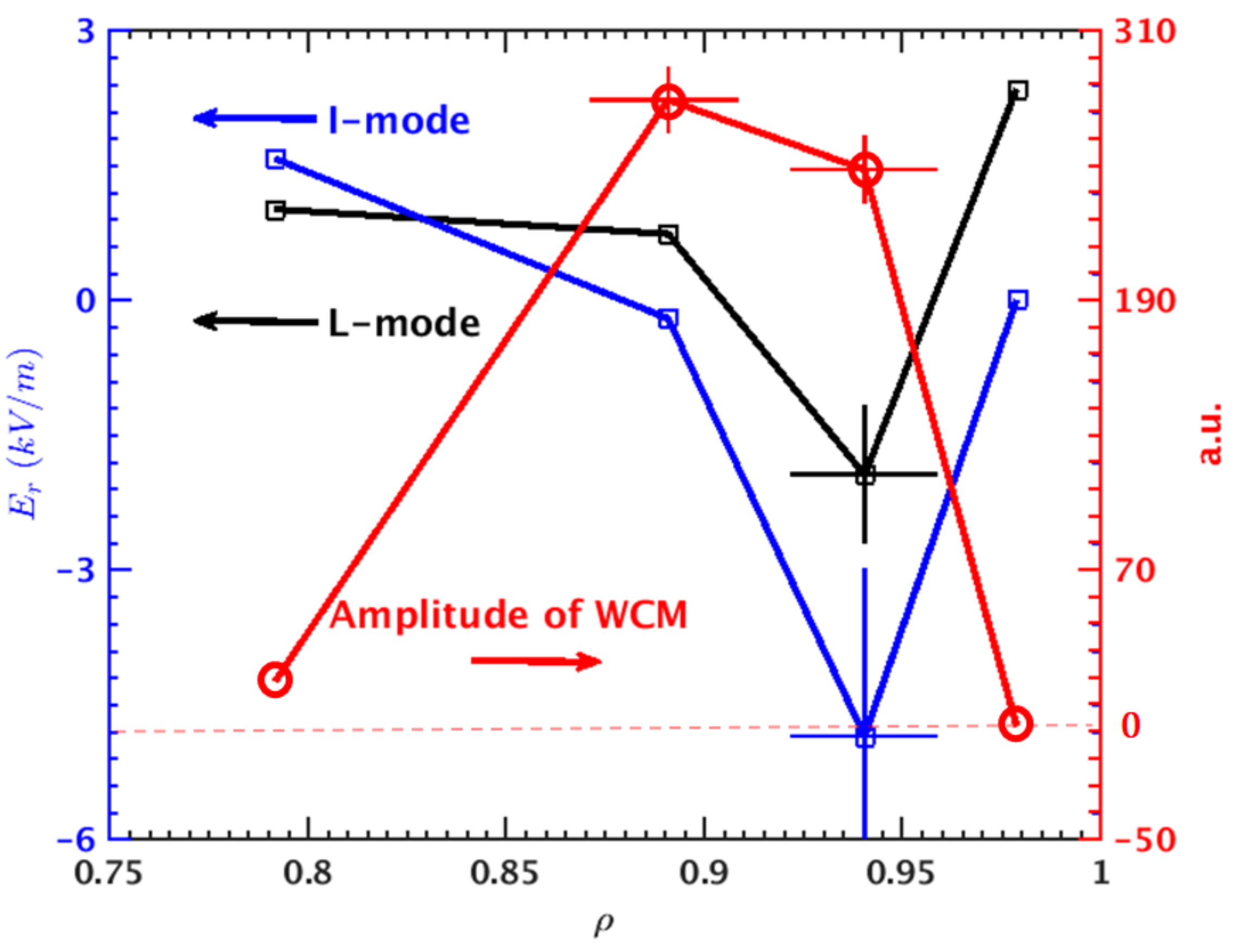}
		\caption{\label{fig:d} Radial distribution of radial electric field $ E_r $ in the L-mode in black and I-mode in blue and the WCM amplitude in red (with background noise subtracted) calculated from the $ E_r $ perturbation spectrum from the DBS8 system of discharge 69967.}
	\end{figure*}
	\begin{figure}
		\includegraphics[width=1\linewidth]{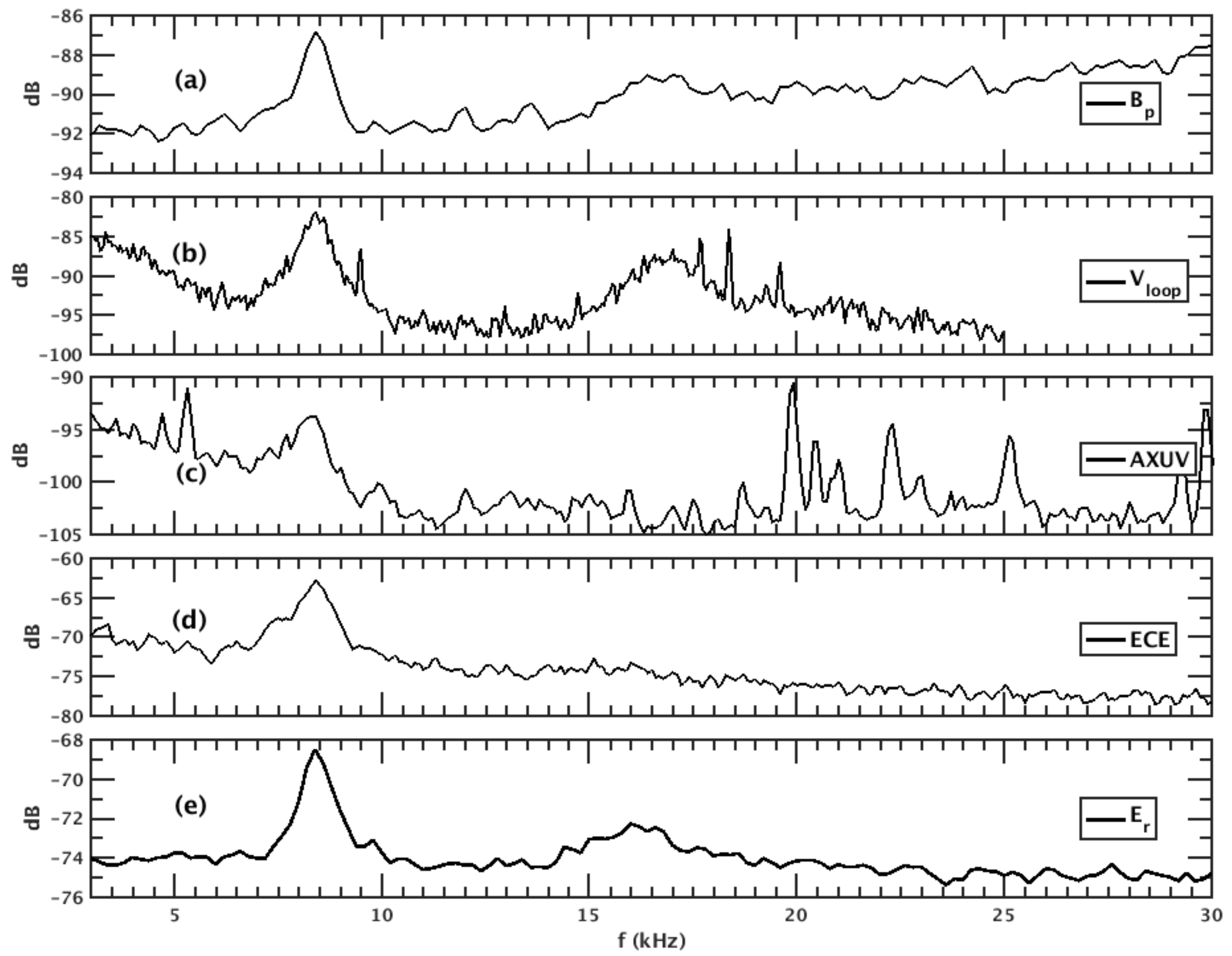}
		\caption{\label{fig:LFO} Examples of the low-frequency oscillation (LFO) at 8.4 kHz in discharge 75280 observed in pick up coil (a), $ V_{loop} $ (b), AXUV (c), ECE (d), and DBS8 (e).}
	\end{figure}
	\begin{figure*}
		\includegraphics[width=1\linewidth]{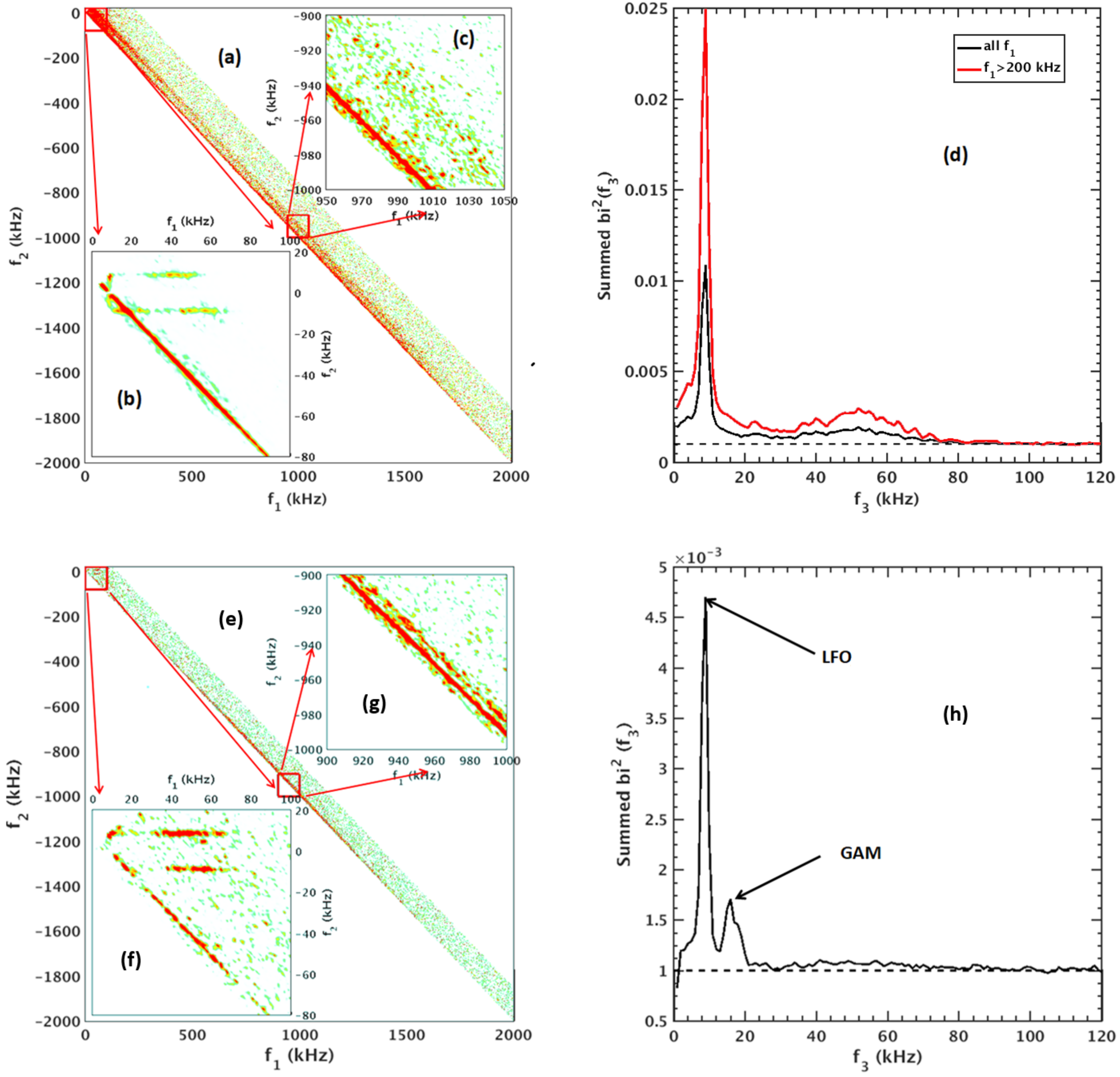}
		\caption{\label{fig:e}	(a) is the cross-bispectrum $ AAd\phi $ in the I-mode region in the DBS channel without GAM in shot 71078, where $ A $ is the density perturbation and d$ \phi $ is the $ E_r $ perturbation. (b) and (c) are enlarged views on low-ferquency ($ f_1\sim  0-100\;kHz$) and high-frequency ($ f_1\sim  950-1050\;kHz$) part, respectively. (d) is the summed bicoherences with the whole frequency range and with only the high frequency range ($ f_1>200\;kHz$). And (e) (f) (g) (h) are the corresponding results with clear GAM in another DBS channel.}
	\end{figure*}
	In EAST, the L-mode to I-mode transition is always accompanied by the WCM with the frequency of 40-150 kHz, similar to the results from other devices. Specially, the WCM in EAST exists in both density fluctuation and $ E_r $ perturbation by modulating the perpendicular flow. Furthermore, the WCM is more prominent in the $ E_r $ fluctuation with a wider radial range without the disturbance from background turbulence. But although the WCM is difficult to see in the density perturbation, it still shows up clearly outside of noise in the power spectrum. However, the WCM is quite weak in magnetic fluctuation in EAST, and because of the low sampling rate of the toroidal magnetic probe, the modulus analysis is currently difficult in EAST.

	A typical example of a perturbation spectrum is shown in Fig.~\ref{fig:c}, and all the perturbation data here are from the channel 67.5 GHz of DBS8 at $ \rho\sim0.97 $. From the ray tracing calculation, the perpendicular wavenumber is approximately $ 4\sim6\;cm^{-1} $. As mentioned above, the low-wavenumber WCM can modulate the cutoff layer; then, it can be observed by the DBS system even at a higher perpendicular wavenumber. In this shot, a clear transition appears at approximately 3.6 s, and the discharge enters a strong I-mode after 3.7 s. Panel (a) shows the time evolution of LHW heating and stored energy $ W_{MHD} $. The evolutions of the density perturbation and radial electric field perturbation spectra are shown in panels (b) and (c), and the evolutions of the intensity of the density fluctuations and the radial electric field ($ E_r $) shear measured by DBS8 are exhibited in panel (d), respectively. Here the intensity of the density fluctuations is the integration of the frequency range 200-700 kHz to avoid the influence of the WCM (the frequency range is about 40-150 kHz). At 3.0 s, along with the increase of the LHW heating power from 300 kW to 1 MW the edge $ E_r $ shear increases dramatically, and the density fluctuation level decreases slightly, but no obvious changes in the density fluctuation spectrum and the $ E_r $ perturbation spectrum are observed. At 3.4 s, the increase of LHW heating power to 2.1 MW increases the edge $ E_r $ shear further, and the density fluctuation level shows a dramatical decrease, then the WCM forms in both density fluctuations and $ E_r $ perturbations, indicating the L-mode to I-mode transition. In the transition process, along with the development of the WCM, the GAM still exists in the $ E_r $ fluctuation, and the frequency increases from 13 to 16 kHz with the increase of the edge temperature. And it should be mentioned that the GAM isn't always observed at the location where the WCM is most significant in I-mode operation, and it may disappear in nearly 60\% I-mode discharges at this location as shown in Fig.~\ref{fig:a}(f) and series (b), (c), (d) of Fig.~\ref{fig:f}, but it may still exist at other radial locations. Simultaneously, the WCM is more clear in the power spectrum as shown in Fig.~\ref{fig:c}(e) and (f), in which the comparison between L-mode (2.6-3 s) and I-mode (3.8-4.2 s) is exhibited, and the peak at approximately 30-90 kHz is distinct both in the density and $ E_r $ fluctuations in the I-mode.

	The comparison of $ E_r $ profiles in the L-mode and I-mode and the radial distribution of the WCM amplitude are shown in Fig.~\ref{fig:d}. For the typical I-mode parameter with $ B_T=2.3\; T $ and the line-averaged density $\bar{n}_e>2.9*10^{19}\; m^{-3} $, only four high-frequency channels of the DBS8 system  inside the separatrix are shown here. In the I-mode, the edge $ E_r $ well is much deeper than that in the L-mode at approximately $ \rho\sim0.94 $, which is consistent with the reports in Alcator C-Mod\cite{R.M.McDermott,C.Theiler} and AUG.\cite{E.Viezzer,F.Ryter} The red line in Fig.~\ref{fig:d} is the radial distribution of the WCM amplitude. And the WCM could only be distinctly observed in the 70 GHz and 72.5 GHz channels and be reluctantly found in the 75 GHz channel, suggesting that WCM locally exists inside the $ E_r $ well and only covers several centimeters ($ 2\sim3 $ cm) at the edge plasmas.

	Specially, a low-frequency oscillation (LFO) with a frequency range of 5-10 kHz always accompanies the WCM in the EAST I-mode operation as shown in Fig.~\ref{fig:a}(f) (approximately 7 kHz) and in Fig.~\ref{fig:c}(a) and (b) (approximately 9 kHz). Many observations show that the LFO is another character of the I-mode operation beside the WCM in EAST, and it seems that LFO always instead of GAM at the position where WCM is generated. Unlike the GAM, the density perturbation and electron temperature perturbation of LFO are both distinct as well as the $ E_r $ perturbation. Actually, the oscillation could be detected by nearly all diagnostics at the plasma edge, such as the magnetic pick-up coils, $ D_\alpha $ filterscopes, divertor probes, edge (ECE) channel, absolute extreme ultra violet diodes (AXUV), soft-X rays, and even loop voltage coils. The power spectra of some signals during I-mode in shot 75280 are shown in Fig.~\ref{fig:LFO}, and these distinct peaks suggest that LFO has large influence on the plasma edge region although its radial extend range is similar to WCM. It should be noted that the second harmonic of LFO could also sometimes be seen, as shown in Fig.~\ref{fig:LFO}(a)(b) and (e). The toroidal symmetry feature of LFO could be easily confirmed from the magnetic coils (not shown), together with the fact that LFO always instead of GAM at the position where WCM is generated, implying that LFO is a novel kind of zonal flows during I-mode. However, there are several significant differences between LFO here and the quasiperiodic $ E_r $ oscillation (also called limited-cycle oscillation, LCO, or dithering sometimes) appearing during L-H transition in EAST and other devices	\cite{G.S.Xu,L.Schmitz,T.Estrada,ChengJ}. Firstly, LFO seems not purely electrostatic, as shown in Fig.~\ref{fig:LFO}(a) and (b); secondly, the typical LFO frequency is about 3 times larger than the LCO in EAST L-H transition (2-3 kHz)\cite{G.S.Xu}, against to the ion-ion collisional damping process of zonal flows; thirdly, the harmonics of zonal flow have never been reported before and is difficult to explain theoretically\cite{P.H.Diamond}. There must be some other process, maybe like the coupling between LFO and WCM, need to be considered in the predator-prey model\cite{K.Miki,Eun-jinKim} to explain these differences. More investigation are needed to draw definitive statements on the nature of LFO.

	To further investigate the interaction among the WCM, LFO and background turbulence, bicoherence analyses are used in this article. The cross-bicoherence of two signals $ \phi_1(t) $ and $ \phi_2(t) $ is defined as:
	\begin{equation}
	b^2(f_1,f_2)=\frac{\left\langle\left|\phi_1(f_1)\phi_1(f_2)\phi_2^*(f_3=f_1+f_2)\right|^2\right\rangle}{\left\langle\left|\phi_1(f_1)\phi_1(f_2)\right|^2\right\rangle\left\langle\left|\phi_2^*(f_3=f_1+f_2)\right|^2\right\rangle}
	\end{equation}
	where $ \phi_j(f_i) $ is the Fourier transform of $ \phi_j(t) $. Fig.~\ref{fig:e} presents the amplitude, amplitude and $ d\phi $ cross-bicoherence\cite{J.C.Hillesheim} from two channels with and without GAM in the DBS8 during I-mode shot 71078. Panels (a) (b) (c) (d) are the channel with only LFO and panels (e) (f) (g) (h) are the adjacent channel with LFO and GAM coexisting. In panel (a) without GAM, two lines are quite bright in Fig.~\ref{fig:e}(a): $ f_1 + f_2 = f_{LFO} $, and $ f_1 + f_2 = f_{WCM} $, which indicates strong nonlinear interactions among the LFO, WCM and background turbulence. And this is the first observation of the coupling between the WCM and background turbulence. Also the enlargement of the low-frequency part in panel (b) shows a strong interaction between the WCM and LFO with a different colorbar, which matches well with the report in AUG\cite{P.Manz-2} but substitutes the GAM with the LFO which may strongly indicate that the LFO observed in EAST plays a similar role in the process of the I-mode development as the GAM reported in AUG\cite{P.Manz-2}. In the summed bicoherence, the peaks at approximately 9 kHz (LFO) and 35-74 kHz (WCM) are distinct, which strongly exceed the significance level, as shown in Fig.~\ref{fig:e}(d). To clearly show the intensity between LFO/WCM and high frequency turbulence ($ >200 \;kHz$), the summed bicoherence with only high frequency components ($ f_1 >200\;kHz $) is also estimated as red line in panel (d), which is about twice the summed bicoherence with all frequency components, indicating that the averaged coupling intensity between LFO/WCM and high frequency turbulence is actually much larger than that between LFO and WCM. These results imply that although the total density fluctuation level is decreasing during I-mode, the coupling between small scale turbulence and large scale coherence modes such as LFO and WCM still plays an important role, and WCM and LFO may both receive energy from background turbulence. For the channel with both GAM and LFO, two lines are observed in panel (g), refered to $ f_1+f_2=f_{GAM} $ and $ f_1+f_2=f_{LFO} $, respectively. Although WCM still exists in this channel, which could be speculated from the couplings between LFO and WCM in panel (f), no obvious coupling between WCM and background turbulence is observed (the summed bicoherence at $ f_3=30-70\;kHz $ a little above the noise level is due to the contribution from the LFO/GAM and WCM couplings), again implying that the nonlinear three-wave interactions between WCM and turbulence mainly occur at the generation location of WCM.
	
	\section{Summary}
\begin{figure*}
		\includegraphics[width=1\linewidth]{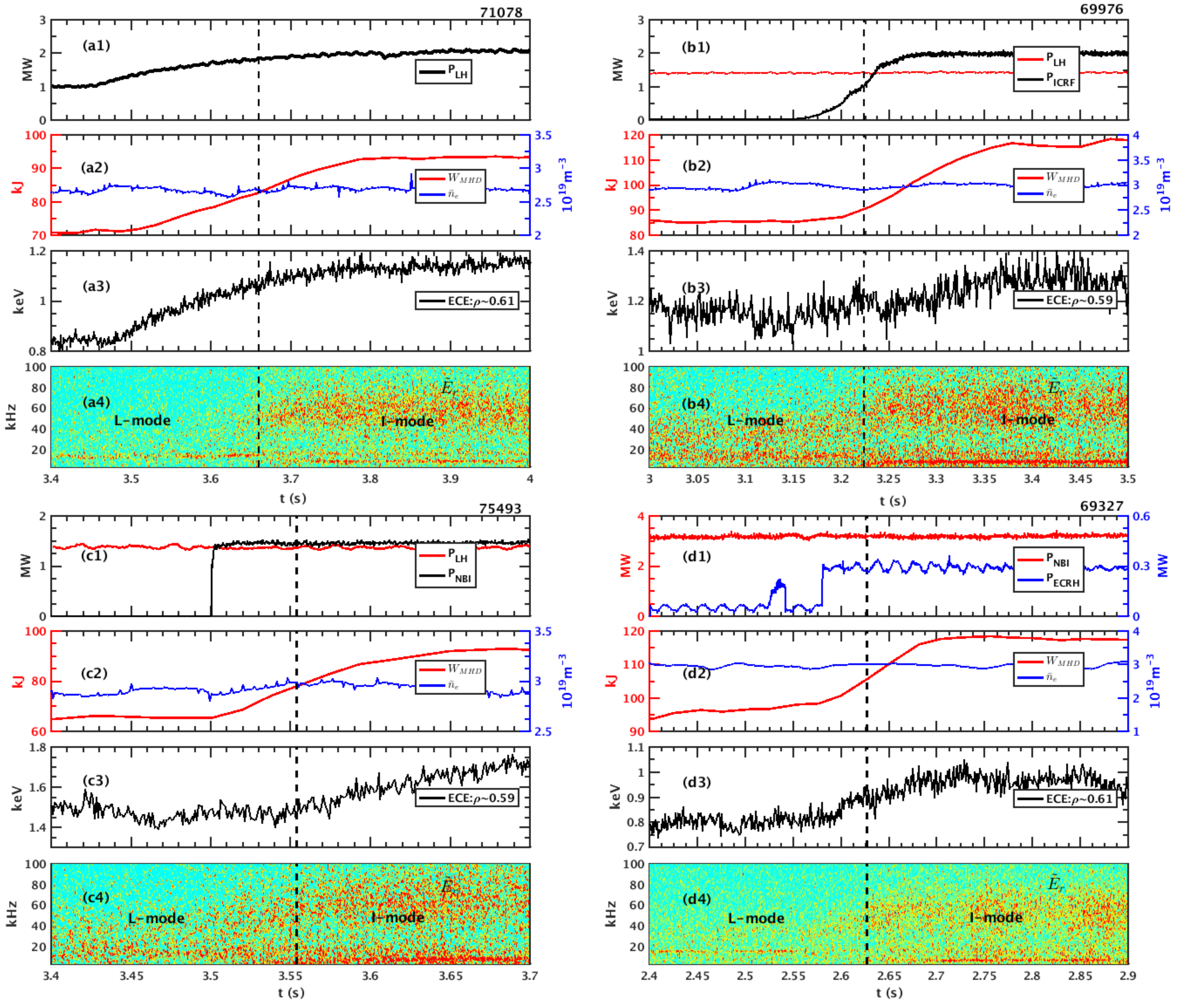}
		\caption{\label{fig:f} Typical I-mode discharges triggered by different auxiliary heating methods. From top to bottom are evolutions of heating power, stored energy, line-averaged density from HCN interferometer, electron temperature from available ECE, radial electric field perturbation from DBS. (a) shot 71078 triggered by LHW only. (b) shot 69976 triggered by ICRF. (c) shot 75493 triggered by NBI. (d) shot 69327 triggered by ECRH.}
	\end{figure*}

	The I-mode operation in EAST has been verified, and it can be maintained for almost the entire flat top in one discharge. The unfavorable ion $ \vec B\times \nabla B $ direction is also a key condition in the I-mode appearance as in Alcator C-Mod, AUG and DIII-D. The L-mode to I-mode transition can be triggered by various heating methods such as LHW heating, NBI, ICRF heating  and ECRH as shown in Fig.~\ref{fig:f}, no heating preference is observed. However, a preference of high plasma currents $ I_p $ is found in EAST. Many observations show that the L-mode to I-mode transition mainly occurs in the discharges with heating power $ >1.8\;MW $ and the density $ >2.5*10^{19}\; m^{-3} $. Generally, the I-mode has a stored energy improvement of nearly 50\% based on the L-mode.

	As a typical characteristic of the I-mode, the WCM is observed in both density perturbation and $ E_r $ perturbation in EAST. An $ E_r $ well is observed at the plasma edge in the L-mode and I-mode regions, and it is deeper in the I-mode, as observed in AUG. The WCM is mainly localized inside position of the $ E_r $ well, and extended radially only 2-3 cm, i.e., could be distinctly observed at one or two channels for DBS diagnostic. Specifically, a low-frequency oscillation with a frequency range of 5-10 kHz is observed to be always accompanied with the WCM in the I-mode operation, Which is regarded as another typical characteristic for I-mode on EAST. It is possible a novel kind of limited-cycle oscillation, but some unique features, like the magnetic components and the harmonics are still need more investigations. And the GAM which is often observed in L-mode region isn't always observed in I-mode operation, and it may disappear in nearly 60\% I-mode discharges at the location where the WCM is the most significant, but it may still exist at other radial locations. Through the bicoherence analysis, it is found that both WCM and LFO strongly couple to the background turbulence in EAST, and interaction between WCM and the LFO also exists strongly, which may indicate that the LFO may play a similar role in the process of the I-mode development as the GAM in AUG, where GAM is found to both modulate and broaden the WCM. And the bicoherence also implys that the coupling between WCM with turbulence is probably a key factor during the generation and saturation processes of WCM. Considering that the intrinsic instability of WCM is still unknown, the role of small scale turbulence should be paid more attention for future research on WCM.

	\ack The authors especially thank Dr. Jerry Hughes for the helpful discussions and Dr. Tianfu Zhou for his kind help in ECE DATA providing. The present work was supported in part by National MCF Energy R\&D Program under Grant No. 2017YE0301204 and 2018YFE0311200, Natural Science Foundation of China under Grant No. 11635008, Anhui Provincial Natural Science Foundation No. 1708085QA23. We also acknowledge the EAST team for the support of these experiments.

\end{document}